\documentclass[12pt]{iopart}
\usepackage[dvips]{graphicx}
\usepackage{amssymb}
\newcommand{\ped}[1]{\ensuremath{_{#1}}}
\newcommand{\apex}[1]{\ensuremath{^{#1}}}
\begin{document}

\title[Point-contact spectroscopy in segregation-free Mg$_{1-x}$Al$_x$B$_{2}$ single crystals]{Point-contact Andreev-reflection spectroscopy in segregation-free Mg$_{1-x}$Al$_x$B$_{2}$ single crystals up to $x=0.32$}

\author{D Daghero\dag, Debora Delaude\dag\, A Calzolari\dag\, M
Tortello\dag\, G A Ummarino\dag\, R S Gonnelli\dag\, V A
Stepanov\ddag\, N D Zhigadlo\S\, S Katrych\S\, and J Karpinski\S }

\address{\dag\ Dipartimento di Fisica and CNISM, Politecnico di
Torino, 10129 Torino, Italy}

\address{\ddag\ P.N. Lebedev Physical Institute, Russian Academy of
Sciences, 119991 Moscow, Russia}

\address{\S\ Laboratory for Solid State Physics, ETHZ, CH-8093
Zurich, Switzerland}

\ead{dario.daghero@polito.it}

\begin{abstract}
We present new results of point-contact Andreev-reflection (PCAR)
spectroscopy in single-phase Mg\ped{1-x}Al\ped{x}B\ped{2} single
crystals with $x$ up to 0.32. Fitting the conductance curves of our
point contacts with the two-band Blonder-Tinkham-Klapwijk model
allowed us to extract the gap amplitudes $\Delta\ped{\sigma}$ and
$\Delta\ped{\pi}$. The gap values agree rather well with other PCAR
results in Al-doped crystals and polycrystals up to $x=0.2$ reported
in literature, and extend them to higher Al contents. In the
low-doping regime, however, we observed an increase in the small gap
$\Delta\ped{\pi}$ on increasing $x$ (or decreasing the local
critical temperature of the junctions, $T\ped{c}^{A}$) which is not
as clearly found in other samples. On further decreasing
$T\ped{c}^{A}$ below 30 K, both the gaps decrease and, up to the
highest doping level $x=0.32$ and down to $T\ped{c}^{A}= 12$ K, no
gap merging is observed. A detailed analysis of the data within the
two-band Eliashberg theory shows that this gap trend can be
explained as being mainly due to the band filling and to an increase
in the interband scattering which is necessary to account for the
increase in $\Delta\ped{\pi}$ at low Al contents ($x<0.1$). We
suggest to interpret the following decrease of $\Delta\ped{\pi}$ for
$T\ped{c}^{A}<30$ K as being governed by the onset of inhomogeneity
and disorder in the Al distribution that partly mask the intrinsic
effects of doping and is not taken into account in standard
theoretical approaches.
\end{abstract}


\submitto{\JPCM}

\maketitle

\section{Introduction}\label{sect:intro}
As well known, the superconductivity in MgB\ped{2} is characterized
by two distinct energy gaps due to the presence of various bands
crossing the Fermi level (generally grouped in two systems: the 3D
$\pi$ bands and the 2D $\sigma$ bands) and to the exceedingly small
quasiparticle scattering between these bands. The values of the gaps
measured in pure MgB\ped{2} by many different techniques agree very
well with those calculated within the two-band models in BCS
\cite{liu01} or Eliashberg approach \cite{choi02}. The multi-band
nature of MgB$_2$ allows explaining most of its features -- in
particular, the relatively high critical temperature and its
unexpected robustness against sample quality -- but also hugely
increases the complexity of the effects that can arise when the
system is in some way ``perturbed''.

A particularly interesting and debated point in the physics of
MgB$_2$ is the possibility to attain the so-called ``gap merging'',
i.e. the complete isotropization of the compound  with consequent
collapse of the two gaps in a single gap with BCS character
\cite{liu01}. Within the two-band Eliashberg theory, the gap merging
can be attained, for example, by keeping all the parameters as in
pure MgB$_2$ and only increasing the interband scattering rate,
$\gamma\ped{\sigma \pi}$. The two gaps approach each other
asymptotically as a function of $\gamma\ped{\sigma \pi}$, while the
critical temperature is reduced; for sufficiently high values of
this parameter ($> 50$ meV), one expects the two gaps to be
virtually indistinguishable. The critical temperature of
``isotropic'' MgB$_2$ varies between 19 and 26 K, depending on the
calculations \cite{choi02,liu01}. Actually, such values of
$\gamma\ped{\sigma \pi}$ are not physical since it is practically
impossible to increase the interband scattering without affecting
other parameters of the material, namely the partial DOS of the
$\sigma$ or $\pi$ bands. This is true for chemical substitutions in
MgB$_2$, e.g. carbon in the B site and aluminum in the Mg site
\cite{cava03}, but also, unexpectedly, for neutron irradiation
\cite{daghero06c}. Generally speaking, the coexistence of various
effects makes it difficult to experimentally single out their
contributions. Fortunately, pair breaking from interband scattering
gives rise to peculiar effects so that it can be separated rather
easily from other sources of $T\ped{c}$ reduction:
$\gamma\ped{\sigma \pi}$ suppresses $T\ped{c}$ and the large gap
$\Delta\ped{\sigma}$, simultaneously increasing the smaller gap
$\Delta\ped{\pi}$.

In a theoretical paper by Erwin and Mazin \cite{erwin03}, Al
substitution in the Mg site was proposed as an effective way to
increase the scattering between bands. First-principle calculations
gave, for 2\% of Al, a value of $\gamma\ped{\sigma \pi}=1.1$ meV,
which is already expected to have measurable effects on the critical
temperature and on the gaps \cite{erwin03}. This made the
Mg$\ped{1-x}$Al$\ped{x}$B$_2$ system the most likely candidate for
the attainment of the gap merging.

In the past years, many experimental efforts have been made to test
these predictions. Aluminum substitution was indeed one of the first
successfully obtained in MgB$_2$ \cite{cava03}. Al atoms in MgB$_2$
are almost completely ionized, exactly like the Mg atoms they
substitute. Aluminum is thus a donor and the three electrons in its
outer shell enter the system of bands giving rise to electron
doping. The effects of the substitution on the lattice are rather
complicated by a strong tendency to the formation of different
phases. Early reports \cite{cava03} showed the presence of two
phases with AlB$_2$ structure and different $c$ axis in
polycrystalline samples of nominal Al content between 0.1 and 0.25.
A similar result was found in single crystals grown at ETH (Zurich)
by means of a high-pressure, cubic-anvil technique. In this case,
the precipitation of a non-superconducting MgAlB$_4$ phase was
observed for $x>0.1$ by High-Resolution TEM and other structural
analysis techniques \cite{karpinski05}. An independent, indirect
confirmation to this picture came from point-contact
Andreev-reflection (PCAR) measurements we performed in those single
crystals, which showed an anomalous trend of the gaps
$\Delta\ped{\sigma}$ and $\Delta\ped{\pi}$ as a function of the Al
content \cite{daghero05,karpinski05}, with a crossover between two
regimes around $x=0.1$. For $x<0.1$, the large gap decreased
linearly with $x$ while the small gap showed a pronounced tendency
to increase, as theoretically expected \cite{erwin03}. For $x>0.1$,
$\Delta\ped{\sigma}$ was found to saturate at about 4 meV, while
$\Delta\ped{\pi}$ was fast suppressed becoming smaller than 1 meV at
$x=0.2$. An analysis of the data within the two-band Eliashberg
theory showed that the trend observed for $x>0.1$ could be explained
by a decrease in the $\pi$-band superconducting coupling, that we
argued could be related to the precipitation of the spurious phase
\cite{karpinski05}.

These anomalies were not confirmed by successive measurements we
performed in polycrystalline Mg\ped{1-x}Al\ped{x}B$_2$ samples grown
in Genova \cite{putti05} that did not suffer from extended phase
segregation even at high Al contents ($x=0.2$). In fact, their
growth technique involved a very long high-temperature reaction (100
h at 1000 $^{\circ}$C) and no evidence of spurious phases was found
by XRD -- even though successive microprobe analyses
(wavelength-dispersive X-ray spectroscopy, WDX) showed a small
amount (4\% of the volume at most) of a secondary phase in the form
of micrometer-size islands embedded in a (Mg,Al)B$\ped{2}$ matrix
\cite{birajdar05} that were concluded to have negligible effect on
the superconducting properties. PCAR measurements we carried out in
these polycrystals showed an almost linear decrease of both
$\Delta\ped{\sigma}$ and $\Delta\ped{\pi}$ as a function of the Al
content which turned out to be in qualitative agreement with the
findings of specific-heat measurements in the same samples
\cite{putti05} as well as with the results of PCAR measurements in
crystals and polycrystals carried out by other groups
\cite{klein06,szabo06}. In all these cases, no gap merging was
observed up to $x=0.2$, but its occurrence at a higher Al content
(corresponding to $T\ped{c}$ around 12 K)\cite{klein06} was
apparently suggested by the overall gap trends.

Recently, new Mg\ped{1-x}Al\ped{x}B\ped{2} crystals have been grown
at ETH that do not show phase segregation up to $x=0.32$. In this
paper we present the results of PCAR measurements on this new
generation of single crystals, and compare them to the results of
PCAR on polycrystals grown in Genova \cite{putti05} as well as to
other data in literature \cite{klein06,szabo06,cooley05}. We will
show that: i) our data extend previous results in single crystals
\cite{klein06} up to the region of extremely high doping; ii) our
data differ from most of the results in literature in the low-doping
region, where we observe a much more marked increase in the small
gap $\Delta\ped{\pi}$ on increasing $x$. Once reported as a function
of the critical temperature of the junctions, $T\ped{c}\apex{A}$,
the values of $\Delta\ped{\pi}$ reach a maximum around
$T\ped{c}\apex{A}=30$ K and then start to decrease; iii) this trend
nicely agrees with that observed by specific-heat measurements in
high-quality polycrystals free of compositional gradients
\cite{cooley05}; iv) our data show no gap merging up to $x=0.32$ and
down to $T\ped{c}\apex{A}=12$ K. We will also show that the trend of
the gaps in our single crystals can be well explained within the
two-band Eliasberg theory as being due to the band filling (which is
the dominant effect of Al doping) and to a substantial increase in
interband scattering in the low-doping region ($x < 0.1$). This
intrinsic effect of Al doping explains the observed initial increase
in $\Delta_{\pi}$ on increasing $x$, while for $x > 0.1$ other
phenomena, e.g. inhomogeneities in the dopant distribution --
witnessed by a sudden increase in the superconducting transition
width -- may concur in making the gaps decrease again. In this range
of doping, all the theoretical models that do not take into account
inhomogeneity and disorder should be used with some caution as their
predictions might not reflect actual properties of the compound.

\section{Experimental}
\subsection{The samples} \label{sub:samples}
\begin{figure}[tb]
\begin{center}
\includegraphics[keepaspectratio, width=0.6\columnwidth]{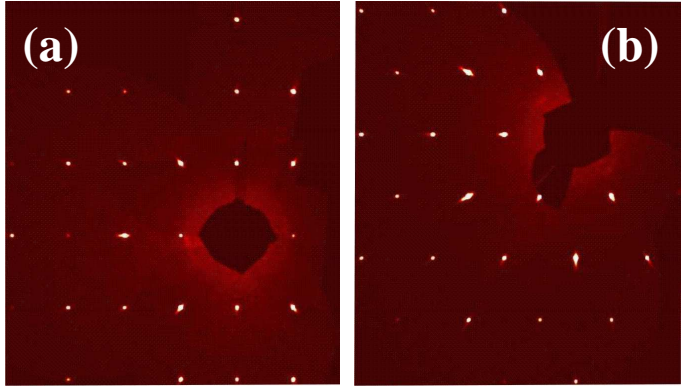}
\end{center}
\caption{\label{fig:Xray}(a) The reconstructed $h0l$ reciprocal
space section of the AN394/5 sample ($x=0.24$). (b) The
reconstructed $hk0$ reciprocal space section of the AN412/5 sample
($x=0.32$).}
\end{figure}

Mg\ped{1-x}Al\ped{x}B\ped{2} single crystals were grown by using the
high-pressure cubic-anvil technique described elsewhere
\cite{karpinski03a,karpinski03b}, starting from pure B powder and a
Mg-Al alloy and tuning time, pressure and temperature to eliminate
phase segregation. As usual, the Al content was determined by energy
dispersive X-ray analysis (EDX) and ranged between 0.02 and 0.32. A
thorough characterization of the structural, morphological, chemical
and superconducting properties was performed, by using the
techniques described in \cite{karpinski05}. In particular, the
quality of the crystals was checked by XRD using a X-ray
diffractometer equipped with CCD area detector (Xcalibur PX, Oxford
Diffraction), which allowed us to examine the whole reciprocal space
(Ewald sphere) for the presence of other phases or crystallites with
different orientation. No additional phases (impurities, twins or
intergrowing crystals) were detected by examination of the
reconstructed reciprocal space sections. This is clearly seen by
comparing \fref{fig:Xray} of this paper to figure 5 of
\cite{karpinski05}.

%
The crystals revealed MgB\ped{2} structure \cite{jones54}. The
structure refinement results of two samples with high Al content are
presented in Table \ref{Table:1}. Because Al and Mg have almost the
same amount of electrons (12 and 13 respectively) the refinement was
performed without Al and the position of Mg was considered to be
occupied with both atoms. The presented refinement results of two
highly-doped samples and their reconstructed reciprocal space
sections show that the crystals chosen for subsequent PCAR analysis
are of a high quality and satisfy the requirements for single
crystals. On the basis of these data, we can exclude the influence
on measured properties of some possible factors like
polycrystallinity, additional phases, strong disorder etc.

\begin{table}
\caption{\label{Table:1}Structure refinement and crystal data for
Al-doped MgB\ped{2} samples AN394/5 and AN412/5.}
\tiny{
\begin{tabular}{@{}llll} \br
&AN394/5& &AN412/5\\
\mr
Empirical formula & (MgAl)\ped{0.94}B\ped{2} & &(MgAl)B\ped{2}\\
Temperature, K  &    & 295(2) &  \\
Wavelength,{\AA}/radiation &  & 0.71073/ \textit{Mo}K$\alpha$ & \\
Diffractometer & & Oxford diffraction 4-circle & \\
& &diffractometer (CCD detector)& \\
Crystal system & & Hexagonal & \\
Space group & & P6/mmm & \\
Unit cell size ({\AA}) & $a=3.0787(4)$ & & $a=3.0673(5)$ \\
& $c=3.5198(4)$& & $c=3.4258(5)$ \\
Unit cell volume ({\AA}$^3$) & 28.892(6) &  & 27.913(8) \\
$Z$ & & 1 & \\
Absorption correction type & & analytical & \\
Crystal size (mm) & $0.33\times 0.15 \times 0.03$ & & $0.31 \times
0.22 \times 0.02$ \\
Theta range (deg) & 5.79 to 36.13 & & 7.69 to 33.05 \\
Limiting indices & $-5\le h \le 4$ & & $-4\le h \le 3$ \\
& $-3\le k \le 4$ & & $-4\le k \le 2$ \\
& $-5\le l \le 5$ & & $-2\le l \le 5$ \\
Reflections collected/unique & 204/46, Rint=0.0364 & & 156/37,
Rint=0.0179 \\
Refinement method &  & Full-matrix least-squares on $F^2$ & \\
Data/restraints/parameters & 46/0/6 & & 37/0/6 \\
Goodness of fit on $F^2$ & 1.005 & & 1.071 \\
Final R indices [$l > 2\sigma(l)$]& R\ped{1}=0.0243,
wR\ped{2}=0.0611 & & R\ped{1}=0.0162, wR\ped{2}=0.0409 \\
R indices (all data) & R\ped{1}=0.0324, wR\ped{2}=0.0620 & &
R\ped{1}=0.0174, wR\ped{2}=0.0411 \\
$\Delta \rho \ped{max},\Delta \rho \ped{min}$ (e/{\AA}$^3$) & 0.354
and -0.350 & & 0.222 and -0.247 \\
\mr
&Fractional atomic coordinates & and atomic displacement parameters ({\AA}$^2$) &\\
\mr
B &  & x=1/3; y=2/3; z=1/2 & \\
(Mg,Al) & & x=0; y=0; z=0 & \\
$U_{11}$, B & 0.010(1) & & 0.007(1)\\
$U_{33}$, B & 0.014(1) & & 0.010(1) \\
$U_{12}$, B & 0.005(1) & & 0.003(1) \\
$U_{11}$, (Mg,Al) & 0.009(1) & & 0.008(1) \\
$U_{33}$, (Mg,Al) & 0.012(1) & & 0.008(1) \\
$U_{12}$, (Mg,Al) & 0.005(1) & & 0.004(1) \\
\br
\end{tabular}}
\end{table}

\begin{figure}
\begin{center}
\includegraphics[keepaspectratio, width=0.6\columnwidth]{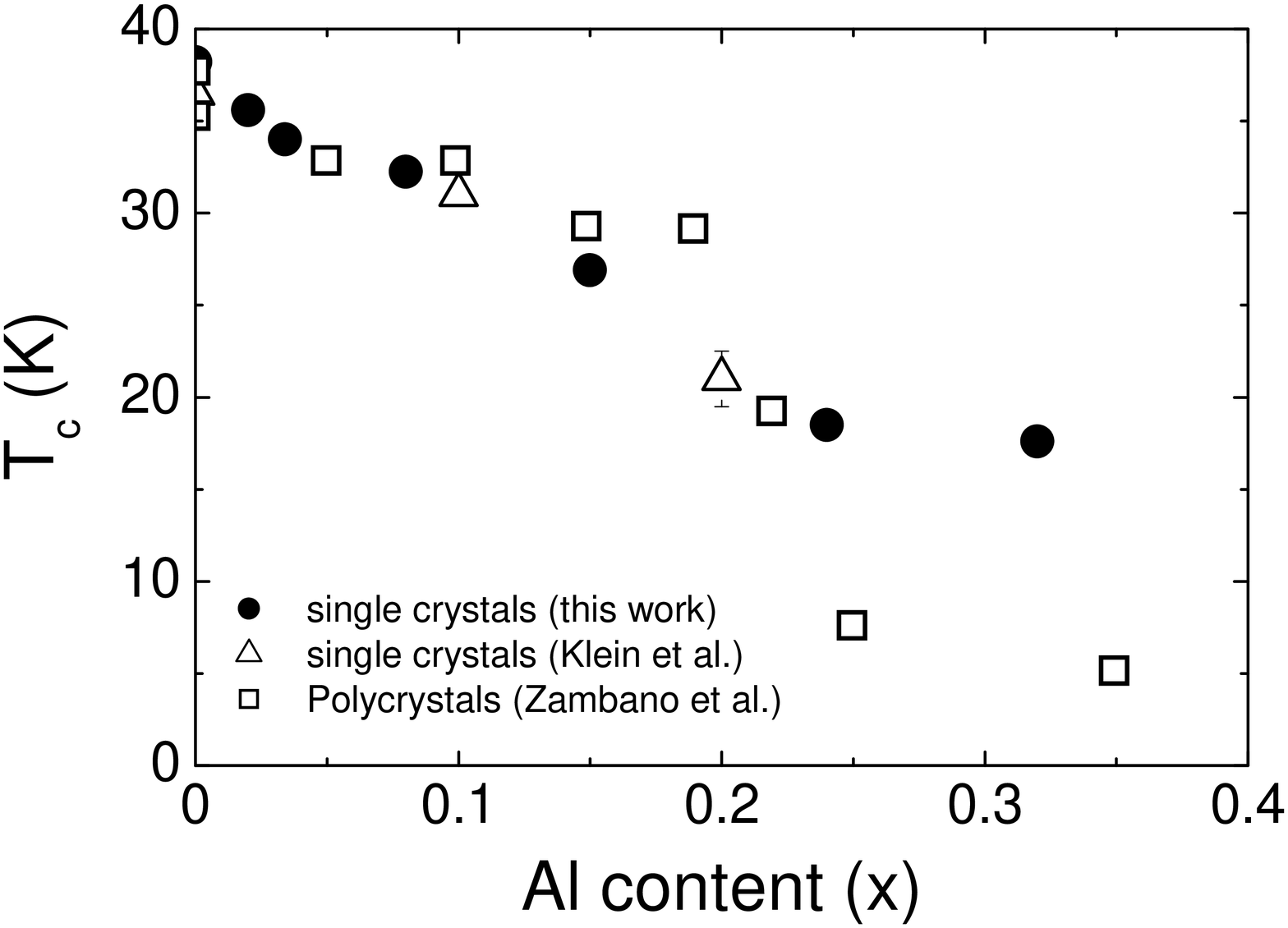}
\end{center}
\caption{\label{fig:tcvsx}Bulk critical temperature $T\ped{c}$
measured as a function of the aluminum content $x$. \fullcircle:
Data obtained in our single crystals by DC susceptibility
\cite{karpinski05}. \opentriangle: Data from \cite{klein06},
obtained in single crystals by specific-heat measurements.
\opensquare: Data from DC susceptibility in long-annealed
polycrystals \cite{zambano05}.}
\end{figure}

Figure~\ref{fig:tcvsx} shows the relation between the Al content and
the bulk T$_{c}$ given by DC susceptibility in our single crystals
(black circles). The critical temperature was defined here as the
abscissa of the intersection between the $y=0$ axis and the linear
fit of the susceptibility vs. $T$ in the region of the transition.
The transition width $\delta T\ped{c}$ (defined as
$T\ped{10\%}-T\ped{90\%}$) increases on increasing the Al content,
from 0.77 K in the crystal with $x=0.02$ up to about 9 K in the most
doped sample ($x=0.32$). The large values of $\delta T\ped{c}$ in
heavily doped samples can be related to the local inhomogeneity in
the dopant content \cite{birajdar05,lee04}, and to the smallness of
the coherence length $\xi$ that allows inhomogeneities on a scale of
the order of $\xi$ to be resolved \cite{zwicknagl84}. A contribution
from the simple disorder consequent to Al doping may be present as
well, but other sources of broadening are either negligible or
excluded by the single-crystal nature of our samples. For instance,
due to the small DC field used in the magnetization measurements (2
to 5 Oe), effects related to the magnetic field are certainly small.
It will be clear in the following that the inhomogeneous
distribution of Al in the crystals does not invalidate the results
of local measurements such as PCAR spectroscopy, provided that the
gap amplitudes are reported as a function of the local critical
temperature in the region of the contact and not as a function of
the (average) Al content or of the bulk $T\ped{c}$.

\subsection{Point-contact Andreev-reflection measurements}\label{sub:PCS}
Before starting with the point-contact measurements, we cleaned the
crystals and etched their surface by dipping them into a solution of
1\% HCl in dry ethanol. After 2-5 minutes, we rinsed the crystals in
pure ethanol and dried them with nitrogen. The point-contact
measurements were performed by using the ``soft'' technique
described elsewhere \cite{gonnelli02c}. Instead of pressing a
metallic tip against the sample as in standard PCAR, we made the
contact by using as a counterelectrode a small spot of Ag conductive
paint. This pressure-less technique can be used also on brittle
samples and, in thin single crystals, allows injecting the current
(mainly) along the $ab$ planes \cite{gonnelli02c}, so as to measure
both the $\sigma$ and $\pi$-band gap at the same time
\cite{brinkman02}. The diameter of the Ag-paint spot is typically
$\varnothing \leq 50\,\mu$m that, however, does not correspond to
the actual size of the point contacts. As a matter of fact, parallel
microjunctions are very likely to form between the crystal surface
and the Ag particles in the paint within the macroscopic contact
area, so that the measured $I-V$ characteristics and conductance
curves should be regarded as an average over a certain region in
direct space. Usually, the potential barrier at the N/S interface is
rather low so that the contacts are in the Andreev-reflection
regime. Otherwise, the characteristics of the contact and its
normal-state resistance can be tuned by using short voltage or
current pulses \cite{daghero06c,gonnelli02c}. The formation or
modification of contacts with the help of electric pulses is well
known in standard electrotechnics \cite{holm58}. It was also used,
already in the Seventies, to create point contacts for phonon
spectroscopy in normal metals \cite{yanson74} or high-quality
Josephson contacts between two superconductors \cite{weitz78}.
During a voltage pulse (in our case of the order of 1 V for some
milliseconds), the contact region can be heated well above the bath
temperature. This phenomenon was shown to give rise to local
annealing in heavily neutron-irradiated MgB$_2$ \cite{daghero06c},
but in the present case this drawback can be ruled out because the
crystals are already well annealed \cite{karpinski05} and no
enhancement of the local $T\ped{c}$ above the bulk value has ever
been observed.

\begin{figure}[tb]
\begin{center}
\includegraphics[keepaspectratio, width=0.5\columnwidth]{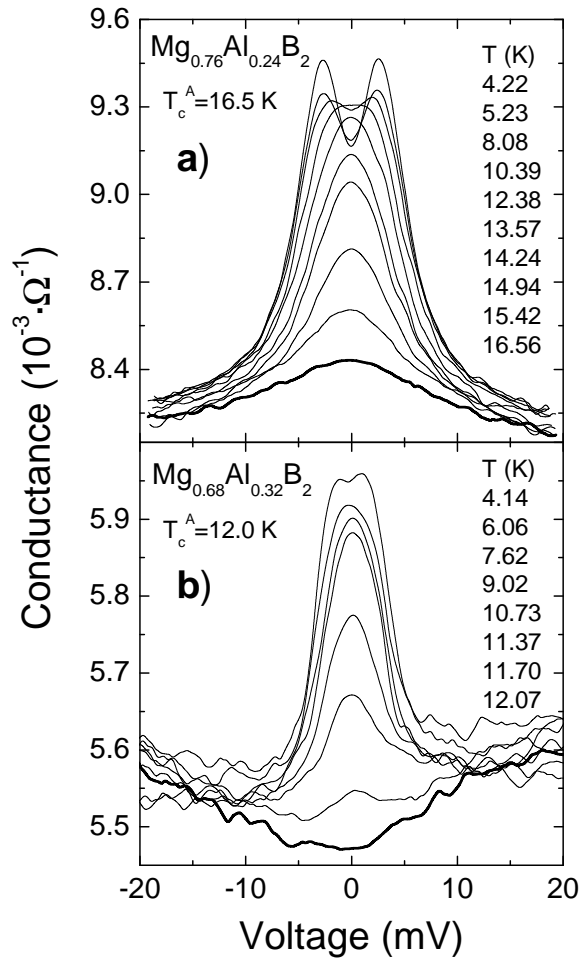}
\end{center}
\caption{\label{fig:Tdep_single}Temperature dependence of two
conductance curves measured in point contacts on crystals with
$x=0.24$ (a) and $x=0.32$ (b). A subset of the complete series is
shown for clarity. The temperatures are indicated in the labels. The
bottom curve in each panel is the normal-state conductance and the
temperature at which it is reached is defined as the Andreev
critical temperature $T\ped{c}\apex{A}$.}
\end{figure}
%
Figure~\ref{fig:Tdep_single} shows two examples of  raw conductance
curves measured in the crystals with $x=0.24$ and $x=0.32$ as a
function of temperature. A subset of the complete series is shown
for clarity. The values of the normal-state resistance of the
junctions are $R\ped{N}=120\, \Omega$ and $R\ped{N}=178\, \Omega$,
respectively. Such high values of $R\ped{N}$ are necessary to
fulfill the conditions for ballistic transport through the junction
\cite{duif89}, because of the shortening of the mean free path due
to Al doping. Let's suppose for a moment that a single contact is
established between the crystal and a Ag grain in the paint; then
the whole resistance is due to this single contact and, using the
residual resistivity of the most-doped crystal ($\rho\ped{0} \simeq
5\, \mu\Omega\cdot$cm), the relationship between $\rho\ped{0}$ and
the mean free path $\ell$ reported in \cite{tarantini06}, and the
Sharvin formula \cite{duif89}, one can evaluate the contact radius
$a$ from the contact resistance. If $R\ped{N} \simeq 180\, \Omega$
the result is $a\simeq 11$ {\AA} that is indeed much smaller than
the mean free path $\ell\simeq 100$ {\AA}. This ensures ballistic
transport even if a single contact is established between sample and
counterelectrode. If several contacts are present, as it probably
happens in our case, each of them has higher resistance than the
parallel as a whole and is thus certainly ballistic.

The temperature at which the Andreev-reflection features disappear
and the normal-state conductance is recovered will be in the
following referred to as the local critical temperature of the
contact, or the ``Andreev critical temperature'' $T\ped{c}\apex{A}$.
The values of $T\ped{c}\apex{A}$ are reported in
\fref{fig:Tdep_single} for both the contacts. In doped samples,
because of the local inhomogeneity in the Al content, different
contacts on the same sample can provide different gap amplitudes and
different $T\ped{c}\apex{A}$. All the values of $T\ped{c}\apex{A}$
are included between the onset and the completion of the magnetic
transition, so that in samples with a wide superconducting
transition (especially the most doped ones) values of
$T\ped{c}\apex{A}$ substantially smaller than the bulk $T\ped{c}$
can be obtained. For these reasons, $T\ped{c}\apex{A}$ is more
appropriate than the bulk $T\ped{c}$ to describe the properties of
the contact.
%
\begin{figure}[tb]
\begin{center}
\includegraphics[keepaspectratio, width=0.5\columnwidth]{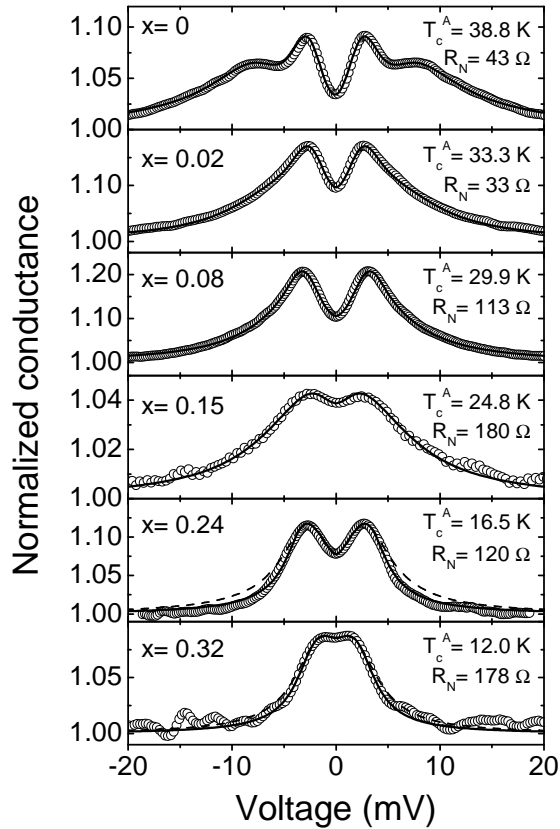}
\end{center}
\caption{\label{fig:totale_single}Normalized low-temperature
conductance curves in crystals with different Al contents from $x=0$
to $x=0.32$. The Andreev critical temperature $T\ped{c}\apex{A}$ and
the normal-state resistance $R\ped{N}$ of each contact are also
indicated. Circles: experimental data. Solid (dashed) lines:
best-fitting curves obtained within the two-band (single band) BTK
model.}
\end{figure}

\section{Results}\label{sect:results}
Figure~\ref{fig:totale_single} shows the conductance curves
$G(V)$=d$I$/d$V$ measured in \textit{ab}-plane contacts in single
crystals with different Al contents $x$ from 0 up to 0.32. All the
curves are normalized, i.e. divided by the normal-state conductance
(measured at $T=T\ped{c}$ or in a magnetic field $H = H\ped{c2}$).
The rather small amplitude of the normalized conductance curves is
related to the ``soft'' point-contact technique we use. In
particular, it is due to some very thin (smaller than the coherence
length $\xi$) impurity layer on the surface of the Ag grains in the
paint, which gives rise to inelastic scattering at the interface. As
shown in \cite{chalsani07}, this effect does not affect the measured
gap values and can be accounted for by simply inserting an extrinsic
broadening in the BTK fit of the conductance curves (see Appendix
A).

To obtain the gap values, we fitted the experimental curves with the
two-band BTK model in which the conductance $G$ of the junction is a
weighed sum of $G_{\sigma}$ and $G_{\pi}$:
$G=w_{\pi}G_{\pi}+(1-w_{\pi})G_{\sigma}$ \cite{brinkman02}.
$G_{\sigma}$ and $G_{\pi}$ depend on the relevant gap amplitude
($\Delta\ped{\sigma}$ or $\Delta\ped{\pi}$), on the effective
potential barrier parameter ($Z\ped{\sigma}$ or $Z\ped{\pi}$), and
on a broadening parameter ($\Gamma\ped{\sigma}$ or
$\Gamma\ped{\pi}$), in conformity to the conventional BTK model
\cite{BTK} modified by including the effect of the quasiparticles
lifetime \cite{plecenik94,srikanth92} (for further details, see
Appendix A). The two-band BTK best-fitting curves are indicated in
\fref{fig:totale_single} as solid lines.

In the low-doping regime (up to $x=0.15$) there is no point to
compare the two-band BTK fit with the single-band one; the latter is
always of poor quality and does not reproduce the position of the
peaks and the width of the Andreev features. However, in the most
doped samples (where, in principle, a gap merging could occur) this
comparison cannot be omitted. As a matter of fact, dashed lines in
the two bottom panels of the same figure represent the single-band
BTK curves that best fit the experimental conductance curves. As
discussed in Appendix A, the two-band fit is always preferable (even
in the $x=0.32$ case, where the two theoretical curves are very
similar to each other) on the basis of a statistical Fisher F-test.
This would lead to the conclusion that \emph{two} gaps are always
present in the Al-doped crystals, even at the highest doping
content.

\begin{figure}[tb]
\begin{center}
\includegraphics[keepaspectratio, width=0.5\columnwidth]{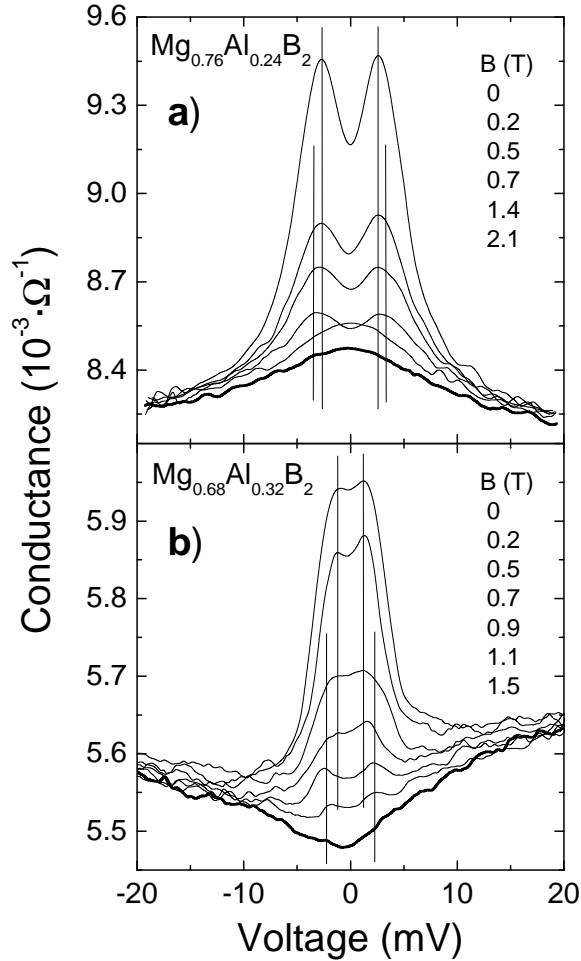}
\end{center}
\caption{\label{fig:Bdep_single}Magnetic-field dependence of the
conductance curves measured on the crystals with $x=0.24$ (a) and
$x=0.32$ (b). The magnetic field was applied parallel to the $ab$
plane. Vertical lines indicate the shift in the position of the
peaks.}
\end{figure}

However, the most reliable test for the actual presence of two gaps
consists in studying the magnetic-field dependence of the
conductance curves, owing to the faster suppression of the
$\pi$-band gap by the magnetic field \cite{gonnelli04c}. In pure
MgB$_2$, this technique allowed us to separate the partial $\sigma$
and $\pi$-band conductances and to fit each of them with a standard,
three-parameter BTK model \cite{gonnelli02c,gonnelli04c}. In doped
samples, the complete separation is not always possible but, if two
gaps are present, an outward shift of the conductance maxima is
observed at a certain magnetic field, when the $\sigma$-band
conductance becomes dominant. Figure~\ref{fig:Bdep_single} reports
the magnetic-field dependence of the conductance curves of the two
contacts in the most Al-doped crystals ($x=0.24$ and $x=0.32$) whose
temperature dependencies have been already shown in
\fref{fig:Tdep_single}. Vertical lines indicate the maximum shift of
the conductance peaks, that witnesses the presence of two gaps
(rather close to each other) and justifies the two-band BTK fit
reported in the two bottom panels of \fref{fig:totale_single}.

\begin{figure}[tb]
\begin{center}
\includegraphics[keepaspectratio, width=0.5\columnwidth]{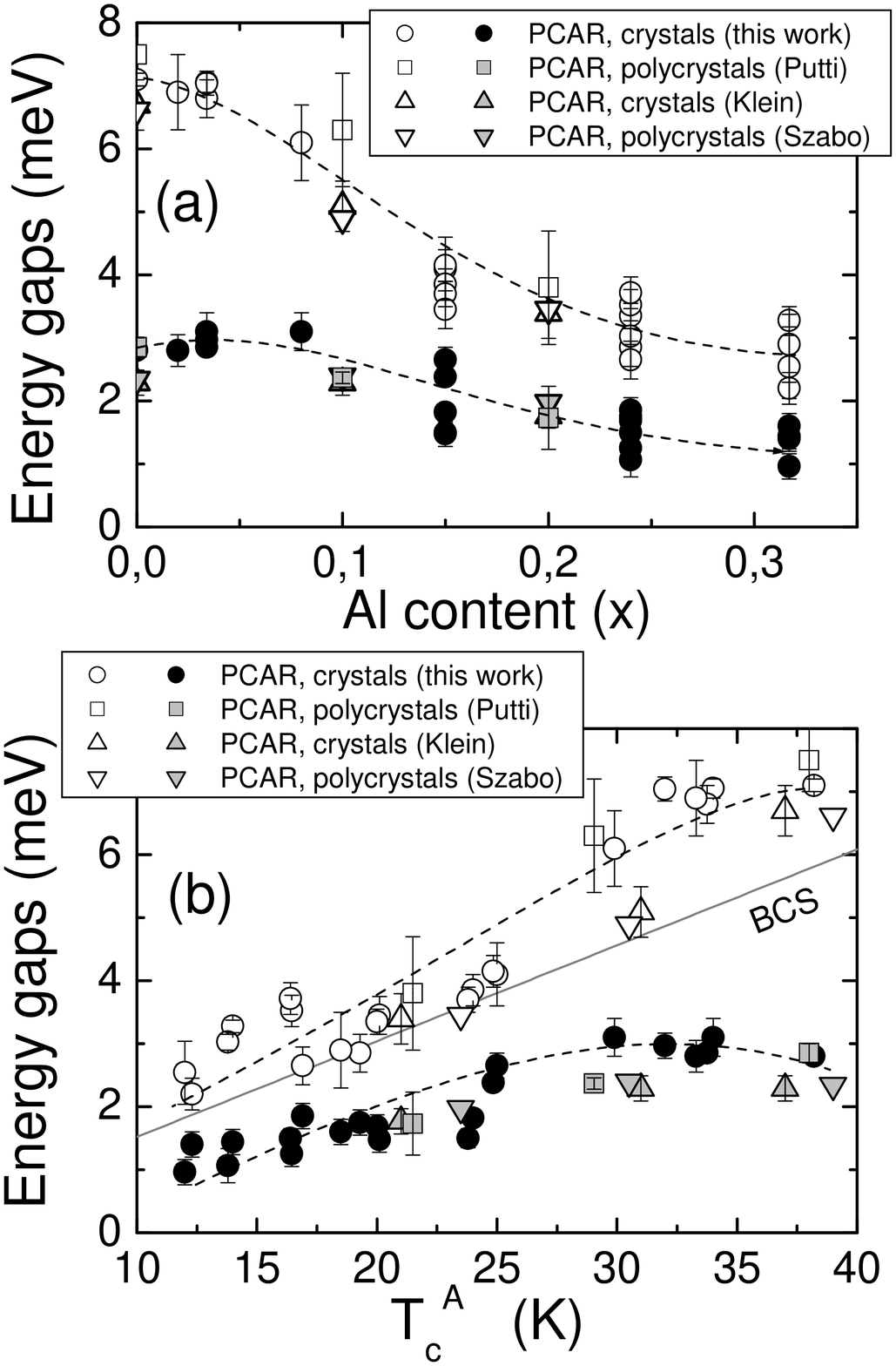}
\end{center}
\caption{(a) Energy gap amplitudes $\Delta\ped{\sigma}$
(\opencircle) and $\Delta\ped{\pi}$ (\fullcircle) measured by PCAR
in single crystals, as a function of the Al content determined by
EDX. Error bars indicate the uncertainty on the gap values for a
given curve. (b) The same values as a function of the critical
temperature of the contact $T\ped{c}\apex{A}$. The gaps (obtained by
PCAR) from refs. \cite{putti05}(squares), \cite{klein06}(triangles)
and \cite{szabo06} (down triangles) are also reported for
comparison. The dashed lines are only guides to the eye, while the
straight grey line in (b) indicates the BCS $\Delta$ vs. $T_c$
dependence.}\label{fig:confronto}
\end{figure}

The gap amplitudes $\Delta\ped{\sigma}$ and $\Delta\ped{\pi}$ given
by the two-band fit of the conductance curves of our point contacts
in Al-doped single crystals (of which \fref{fig:totale_single}
showed a subset) are reported as a function of the Al content $x$ in
\fref{fig:confronto}(a). The vertical spread of data for each doping
content gives an idea of the variation in the local gap values in
different contacts on the same crystal. The trend of the gaps
$\Delta\ped{\sigma}$ and $\Delta\ped{\pi}$ is however clear: The
large gap monotonically decreases on increasing $x$ in the whole
doping range, while the small gap first slightly increases --
reaching a maximum of 3.1 meV at $x=0.08$ -- and then starts to
decrease. For $x>0.15$, the slope of the two curves is apparently
the same.

The vertical dispersion of data can be partly removed by plotting
the gaps as a function of the critical temperature of the contacts,
$T\ped{c}\apex{A}$, as in \fref{fig:confronto}(b). The same figures
also report other PCAR data from literature, obtained by us in
polycrystals \cite{putti05} (squares) and by other groups in single
crystals~\cite{klein06} (up triangles) and polycrystals
\cite{szabo06} (down triangles). Recent STM measurements of the
$\pi$-band gap in crystals grown at ETH substantially agree with
these data \cite{giubileo06}.

In the high-doping region ($T\ped{c}\apex{A} < 25$ K) all the data
sets agree very well with one another independent of the nature of
the samples. In particular, our data nicely extend the curves
previously obtained in single crystals by Klein \emph{et al.} down
to very low critical temperatures (well below the theoretical limit
for gap merging in the hypothesis of pure interband scattering
\cite{liu01,kortus05}). In the low-doping region, a substantially
common trend is observed for the large gap $\Delta\ped{\sigma}$,
although our data (both in crystals and polycrystals) are a little
higher than those reported in literature \cite{klein06,szabo06}. One
may wonder whether this shift is related to the ``soft'' PCAR
technique. This is not the case, since in pure MgB$_2$ we obtained
gap values in excellent agreement with those shown here (at the
highest $T\ped{c}\apex{A}$) also by standard PCAR measurements with
Au or Pt tips \cite{gonnelli02a}. Whatever its origin, this shift
makes our values for $\Delta\ped{\sigma}$ and $\Delta\ped{\pi}$ fall
above and below the BCS value (straight line in
\fref{fig:confronto}b), respectively, while in many data reported in
literature $\Delta\ped{\sigma}$ is very close to the BCS line or
even falls below it. A similar effect has been recently observed in
heavily neutron-irradiated MgB$_2$, but in that case it can be
somehow related to the high level of disorder, on the basis of
analogous evidences for conventional superconductors. Here, this
anomaly is very difficult to justify theoretically -- and is
certainly beyond current models for two-gap superconductivity.

As far as the small gap $\Delta\ped{\pi}$ is concerned, our data
show that, in the low-doping region, it increases on increasing the
Al content, reaching a maximum around $T\ped{c}\apex{A}=30$ K. This
effect is quite definitely assessed by the measurements we expressly
carried out in crystals with $x=0.02$ and 0.034. Such a tendency is
also present in the specific-heat data from \cite{cooley05} and,
although much smaller, in the data from \cite{klein06} and
\cite{szabo06} as well as in the results of specific-heat
measurements in polycrystals from Genova \cite{putti05,putti03}. The
tendency of the small gap to increase was recognized in
\cite{zambano05} as an intrinsic effect of Al doping, more evident
in samples produced via a long reaction at high temperature so as to
reduce the strain and the inhomogeneity in the Al content. The much
greater increase in $\Delta\ped{\pi}$ at low doping contents in the
single crystals grown at ETH (similar to that of samples of the
``B'' series in \cite{zambano05}), might then be related to the
absence of lattice strain due to compositional gradients, also
witnessed by the rather sharp transition of the crystals in this
range of doping levels. If this picture is valid, the increasing
inhomogeneity of the crystals on increasing $x$ might be responsible
for the fact that, for $T\ped{c}\apex{A} < 25$ K, the small gap of
these single crystals returns on the same curve described by the
values of $\Delta\ped{\pi}$ in the other samples.

\begin{figure}[tb]
\begin{center}
\includegraphics[keepaspectratio, width=0.7\columnwidth]{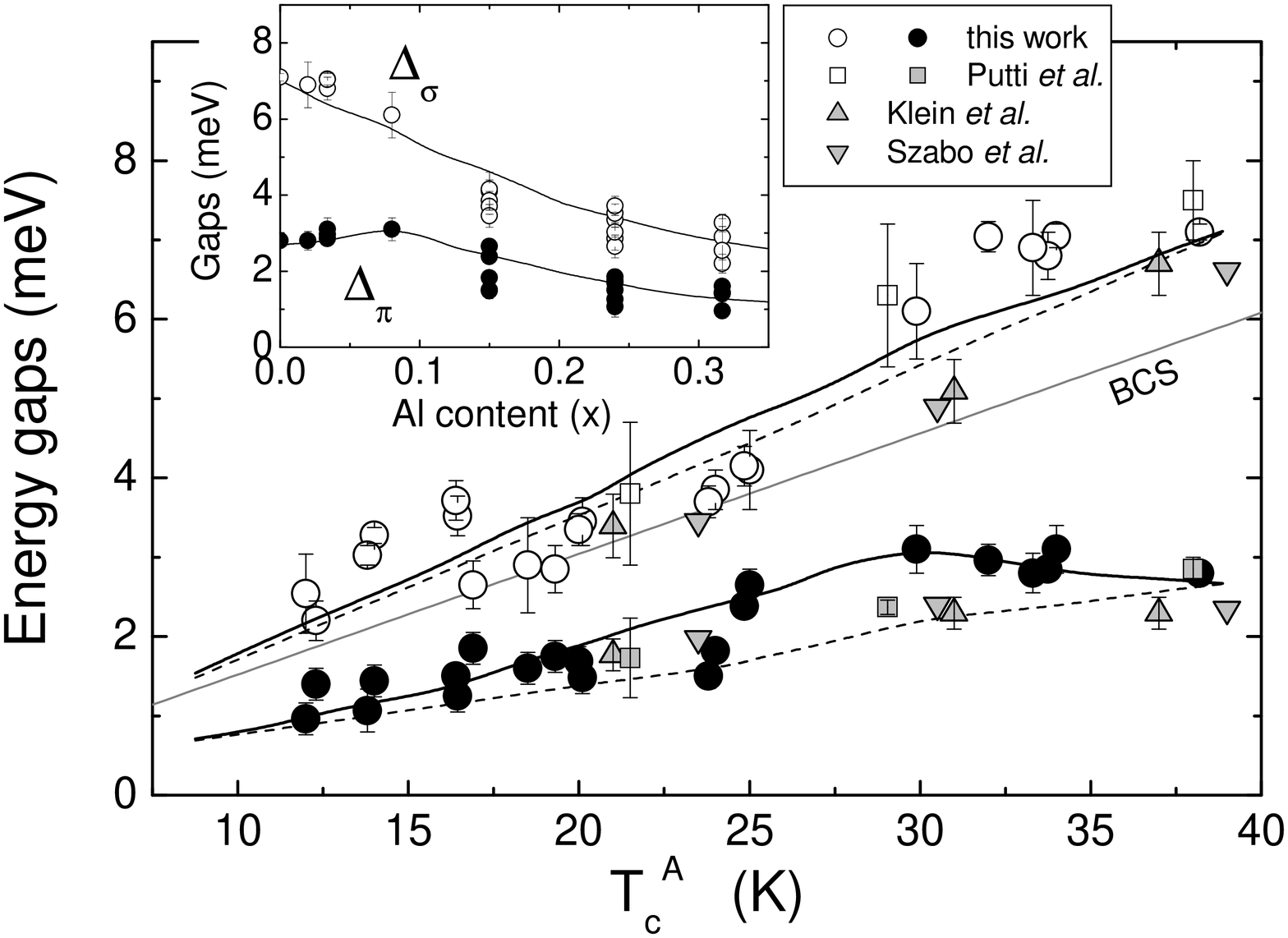}
\end{center}
\caption{\small{Main panel: Energy gap amplitudes in our Al-doped
MgB$_2$ single crystals (\fullcircle,\opencircle) as a function of
the Andreev critical temperature T$_{c}^{A}$. Gap amplitudes from
other PCAR measurements are reported for comparison (squares
\cite{putti05}, up triangles \cite{klein06} and down triangles
\cite{szabo06}). Lines indicate the predictions from Eliashberg
theory, when only the DOSs are changed according to the band filling
due to electron doping (\dashed), and when a proper increase in the
interband scattering at intermediate doping levels is also included
in the model (\full). The straight grey line represents the values
of a BCS gap. Inset: doping dependence of the gaps in our single
crystals, compared to the Eliashberg curves (DOS + interband
scattering). }} \label{fig:Eliashberg}
\end{figure}

\section{Discussion}
We tried to interpret the trend of the gaps as a function of the
local critical temperature within the two-band Eliashberg theory. A
model for the effect of Al doping on the gaps of MgB$_2$ was given
by Kortus et al. \cite{kortus05} who solved the two-band Eliashberg
equations scaling the Eliashberg functions by the change of the DOS
alone. The same approach was used in \cite{ummarino04} to explain
the $x$ dependence of the critical temperature in Al-doped MgB$_2$.
This model has no free parameters as long as one takes the interband
scattering rate to remain negligible as it is in pure MgB$_2$ (see
Appendix B for the explicit Eliashberg equations and the details of
the model). However, even in this case it proved sufficient to
qualitatively explain the previous experimental data for the gaps in
Al-doped MgB$_2$ \cite{putti03,karpinski05} as a function of the
critical temperature, thus indicating that the changes in the
$\sigma$- and $\pi$-band DOS are by far the dominant effect of Al
doping.

Dashed lines in \fref{fig:Eliashberg} represent indeed the
$T\ped{c}\apex{A}$-dependencies of the gaps $\Delta\ped{\sigma}$ and
$\Delta\ped{\pi}$ we calculated within the two-band Eliashberg
theory following the aforementioned approach
\cite{kortus05,ummarino04} and using the DOS N\ped{\sigma}(E\ped{F})
and N\ped{\pi}(E\ped{F}) from first-principle calculations
\cite{profeta03}. It is clearly seen that the dashed lines already
reproduce rather well the data we obtained by PCAR in polycrystals
\cite{putti05} (squares) and those reported in Refs. \cite{klein06}
and \cite{szabo06}. As far as the PCAR data in our single crystals
are concerned (circles in \fref{fig:Eliashberg}), it is clear that
the large gap would be perfectly compatible with the DOS scaling
alone (dashed lines), while the small gap is definitely not. As a
matter of fact, the initial marked increase in $\Delta\ped{\pi}$ on
decreasing $T\ped{c}\apex{A}$ requires some additional ingredient in
the model. According to the discussion of \sref{sect:intro} and to
the predictions of \cite{erwin03}, this immediately suggests an
increase in the interband scattering parameter $\gamma\ped{\sigma
\pi}$ due to the Al doping (incidentally, let's recall that the
intraband scattering parameters, which certainly increase with $x$,
are however ineffective in changing the gap(s), according to
Anderson's theorem. Strictly speaking, this is only true as long as
the doping content is small so that the perturbative description of
the doped compound is possible). By taking $\gamma\ped{\sigma \pi}$
as the only adjustable parameter\footnote{Actually, a very small
change in the prefactor of the Coulomb pseudopotential,
$\mu\ped{0}$, was necessary as well to consistently reproduce the
correct $T\ped{c}\apex{A}$ values. However, $\mu\ped{0}$ varies from
0.031 at $T\ped{c}\apex{A}=39$ K to a minimum of 0.027 reached when
$T\ped{c}\apex{A}=30$ K.} and using the calculated DOS (as we did
for the dashed lines in \fref{fig:Eliashberg}), we were able to
reproduce both the critical temperature and the gap values of our
single crystals. The resulting curves for $\Delta\ped{\sigma}$ and
$\Delta\ped{\pi}$ are shown as solid lines in \fref{fig:Eliashberg}.

To obtain these curves, a non-monotonic dependence of
$\gamma\ped{\sigma \pi}$ on $T\ped{c}\apex{A}$ was necessary.
$\gamma\ped{\sigma \pi}$ initially increases (almost quadratically)
on decreasing $T\ped{c}\apex{A}$, reaching a maximum
$\gamma\ped{\sigma \pi}\apex{max}=2.6$ meV at
$T\ped{c}\apex{A}\simeq 29$ K (i.e., $x\simeq 0.1$), then decreases
linearly to finally saturate to 0.22 meV at $T\ped{c}\apex{A}=15$ K.
At $T\ped{c}\apex{A}=33.5$ K (that means $x\simeq 0.02$), the value
of the interband scattering parameter used for the fit
($\gamma\ped{\sigma \pi}=1.0$ meV) is perfectly compatible with the
theoretical predictions ($\gamma\ped{\sigma \pi}=1.1$
meV)\cite{erwin03}.

If the initial increase in $\gamma\ped{\sigma \pi}$ on increasing
the Al content is easily explained in terms of out-of-plane
distortions of the B sublattice \cite{erwin03}, the problem arises
of explaining the decrease in $\gamma\ped{\sigma \pi}$ at higher Al
contents that is necessary to reproduce the observed gap trend in
single crystals, which shows no gap merging down to
$T\ped{c}\apex{A}$ values as small as 12~K. Giving a definitive
answer to this problem certainly requires further theoretical and
experimental investigations. Here, a simple interpretation can be
anticipated. As discussed above, the comparison of our experimental
data to those of \cite{cooley05} and \cite{zambano05} suggests that
the initial increase in $\Delta\ped{\pi}$, and the corresponding
increase in interband scattering, are intrinsic effects of Al doping
in MgB$_2$, as theoretically predicted. At the end of
\sref{sect:results}, we have also proposed to interpret the decrease
in $\Delta\ped{\pi}$ for $T_{c}^{A} < 30$~K ($x>0.1$) as being due
to the onset of inhomogeneity in the Al content. As a matter of
fact, just in the same region the slope of $\delta T_{c}$ as a
function of $T_{c}$ (being $\delta T_{c}$ the width of the
superconducting magnetic transition) suddenly increases. The
simplified Eliashberg model we have been using (and which is
described in detail in \cite {ummarino04} as well as in Appendix B)
is unsuited to take into account these effects. As a matter of fact,
it is a mean-field model that treats the doped MgB$_2$ as a
perturbation of the pure compound so that: i) the Anderson's theorem
holds; ii) the mathematical expression for the matrix elements of
the Coulomb pseudopotential and of the coupling constant derived for
pure MgB$_2$ holds as well. This may not be true any longer when the
doping concentration is too large: in this case the model itself
probably fails and a different, non-perturbative description should
be used. In this sense, the decrease in $\gamma_{\sigma \pi}$ for
$T_{c}^{A}<30$~K necessary to fit the experimental $\Delta_{\pi}$
values may not reflect an actual property of the system. In other
words, this decrease might be necessary to mimic the effects of
lattice stress and inhomogeneity (possibly at a nanometric scale) in
the local Al content that, at the present moment, are not explicitly
included in the model. Incidentally, this situation is somehow
similar to what we observed in heavily neutron-irradiated MgB$_2$
\cite{daghero06c}.

\section{Conclusions}
In conclusion, we performed a large number of point-contact
Andreev-reflection measurements on segregation-free,
state-of-the-art single crystals of Mg\ped{1-x}Al\ped{x}B\ped{2}
extending previous PCAR results up to $x = 0.32$ and down to local
critical temperatures $T\ped{c}\apex{A} \simeq 12$ K. The local
critical temperature of each contact was directly obtained from the
temperature dependency of the conductance, while the gap amplitudes
were determined by a two-band BTK fit of its low-temperature bias
dependency. No merging of the energy gaps has been observed down to
the lowest $T\ped{c}\apex{A}$ and the persistence of two gaps at the
highest values of $x$ has been confirmed by studying the conductance
curves in the presence of suitable magnetic fields. When compared to
the theoretical results obtained in the framework of the two-band
Eliashberg theory, the experimental gaps show that the main effect
of Al doping is to fill up the bands, thus changing the DOSs at the
Fermi level just as expected by first-principles calculations.
Nevertheless, at an intermediate aluminum content, corresponding to
$T\ped{c}\apex{A}$ values between 18 K and 35 K, the
$\Delta\ped{\pi}(T\ped{c}\apex{A})$ curve of our single crystals
shows clear deviations from the theoretical behaviour expected for
pure band filling. Within the two-band Eliashberg model already used
to describe pure MgB$_2$, these deviations can only be reproduced by
introducing a proper amount (up to 2.6 meV) of interband scattering
$\gamma\ped{\sigma \pi}$. However, $\gamma\ped{\sigma \pi}$ must be
decreased again on further increasing $x$ to account for the
experimental behaviour of $\Delta\ped{\pi}$ and the absence of gap
merging -- at least down to $T\ped{c}\apex{A} =12$ K. We propose to
interpret this trend as resulting from two competing phenomena: i)
the increase in interband scattering, intrinsic to Al doping but
clearly observable only when the lattice stress due to compositional
gradients is eliminated, and ii) the gradual onset of inhomogeneity
when the Al content is increased above a certain threshold ($x\simeq
0.1$). At high doping contents, inhomogeneity probably dominates so
that any theoretical description that does not take it into account
becomes less and less satisfactory. To overcome this problem, which
is likely to occur in any doped MgB$_2$-based system, a detailed
experimental knowledge of the kind of disorder would be required, as
well as new and specific theoretical approaches.

\ack We are indebted to Marina Putti, Andrea Palenzona and Pietro
Manfrinetti for continuous collaboration and very fruitful
discussions. This work was done within the PRIN Project No.
2004022024. V.A.S. acknowledges support by Russian Foundation for
Basic Research (Proj. No. 06-02-16490).

\appendix
\section*{Appendix A. The two-band BTK fit}\label{app:A}
\setcounter{section}{1}

In this Appendix we will give some additional details about the
two-band BTK fit and the fitting procedure. In the theoretical model
we used, the normalized conductance through the point contact is
expressed by:
\begin{equation}
G = w_{\pi}G_{\pi}+ (1-w_{\pi})G_{\sigma}
\end{equation}
Each conductance is expressed in the form

\begin{equation}
G_{i}(E)=\frac{\int_{-\pi/2}^{\pi/2}\sigma_{S,i}(E,\phi)\cos(\phi)\mathrm{d}\phi
}{\int_{-\pi/2}^{\pi/2}\sigma_{N,i}(\phi)\cos(\phi)\mathrm{d}\phi}
\label{eq:conduttanza}
\end{equation}
where $i=\sigma, \pi$ and:
\begin{equation}
\sigma_{N,i}(\phi)=\frac{\cos(\phi)^2}{\cos(\phi)^2+Z_i^2}
\end{equation}

\begin{equation}
\sigma_{S,i}(E, \phi)=\sigma_{N,i}(\phi)\frac{
1+\sigma_{N,i}(\phi)|F_{i}(E)|^2 +
(\sigma_{N,i}(\phi)-1)|F_{i}(E)^2|^2
}{|1+(\sigma_{N,i}(\phi)-1)F_{i}(E)^2 |^2}.
\end{equation}

The functions $F_{i}(E)$ are given by

\begin{equation}
F_{i}(E)=\frac{(E+i\Gamma_{i})-\sqrt{(E+i\Gamma_{i})^2-\Delta_{i}^2}}{|\Delta_{i}|}
\end{equation}
and contain the broadening parameters $\Gamma_{\sigma}$ and
$\Gamma_{\pi}$ as imaginary parts of the energy \cite{dynes78}. Also
note that these parameters are independent of the gap values
$\Delta_i$. This model is a two-band generalization of the
formulation by Kashiwaya \emph{et al.} \cite{kashiwaya96} that
reduces to the simplest BTK formulation if one takes $\Gamma_i=0$
and $\phi=0$ instead of integrating over the angle as in equation
\ref{eq:conduttanza}.

In the modified BTK model \cite{plecenik94,srikanth92} $\Gamma$ is a
measure of the intrinsic lifetime broadening. In our case,
$\Gamma_{\sigma}$ and $\Gamma_{\pi}$ account for both the intrinsic
lifetime broadening and other effects - related to the experimental
technique and thus extrinsic - that smooth the curves
\cite{gonnelli02c}. The most probable origin of the additional
broadening is inelastic quasiparticle scattering in the vicinity of
the contact, i.e. in a degraded layer covering the Ag grains of the
paint. As recently shown \cite{chalsani07} this scattering can be
simply accounted for by increasing the broadening parameter $\Gamma$
in the BTK model. $Z\ped{\sigma}$ and $Z\ped{\pi}$ depend on the
potential barrier at the interface. The weight of the $\pi$-band
conductance, $w\ped{\pi}$, is taken as an adjustable parameter as
well.

The range of variability of the seven fitting parameters is actually
limited by some physical constraints. For example, in pure MgB$_2$
$w\ped{\pi}$ must vary between 0.66 (for pure $ab$-plane tunneling)
and 0.99 (for $c$-axis tunneling) \cite{brinkman02}. Owing to the
non-perfect directionality of PCAR, we always took $0.68 \leq
w\ped{\pi} \leq 0.75$ for $ab$-plane current injection
\cite{gonnelli02c}. Moreover, $w\ped{\pi}$ and the barrier
parameters $Z\ped{\sigma}$ and $Z\ped{\pi}$ must be independent of
temperature and magnetic field so that one is forced to keep them
constant in fitting the whole $T$ or $H$-dependence of a conductance
curve. In principle, $\Gamma\ped{\sigma}$ and $\Gamma\ped{\pi}$
should be smaller than $\Delta_{\sigma}$ and $\Delta_{\pi}$,
respectively, and they must increase on increasing the applied
magnetic field \cite{gonnelli04c}.
\begin{figure}[tb]
\begin{center}
\includegraphics[keepaspectratio, width=0.7\columnwidth]{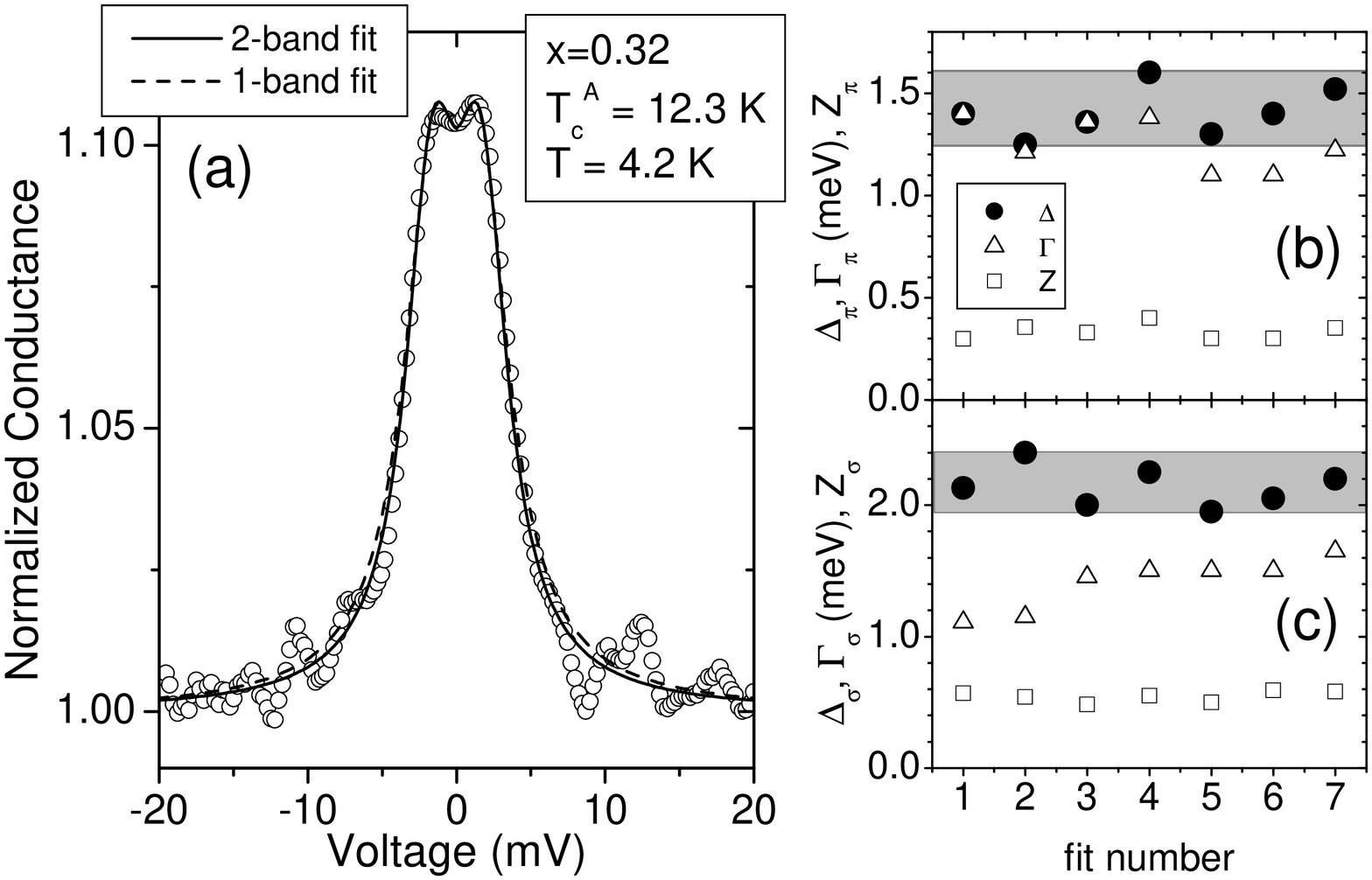}
\end{center}
\caption{\label{fig:parameters}(a): normalized experimental
conductance curve of a point contact in a crystal with $x=0.32$
(\opencircle) compared to the single-band (\dashed) and two-band
(\full) BTK fit. (b) The values of $\Delta$ (\fullcircle), $\Gamma$
(\opentriangle) and $Z$ (\opensquare) for the $\pi$ band, that allow
fitting the same curve, in a series of different fits. (c) The same
as in (b) but for the $\sigma$ band. The variations in the gaps are
qualitatively indicated by grey strips.}
\end{figure}

Finally, the uncertainty on $\Delta_{\pi}$ and $\Delta_{\sigma}$
\emph{for a given curve} can be defined as the maximum range of gap
values that allow a good fit of the curve, when the other parameters
are changed too. To define which is the best fit, we minimized the
sum of squared residuals (SSR). This corresponds to minimizing the
chi-square but does not require an estimation of the uncertainty on
the conductance for each point, which can vary from curve to curve
and is often difficult to estimate. We then allowed a variation of
the SSR of the order of 100\% and determined the corresponding range
of parameters. In the best cases, as in pure MgB$_2$ (top panel of
\fref{fig:totale_single}), $\Delta_{\pi}$ is directly related to the
position of the peaks in the conductance curves, and
$\Delta_{\sigma}$ to the shoulders on the sides. In this situation,
the uncertainty on the gap values is usually rather small (of the
order of 0.3 meV). In doped samples, clear conductance peaks are
still present but no structure at $V>V\ped{peak}$ is directly
visible that would reveal the presence of a second gap. Moreover, in
heavily doped crystals, the two gaps can be so close to each other
that the conductance peaks can occur at some intermediate energy. In
all these cases, a statistical test (the Fisher F-test) can be used
to determine whether the single-band or two-band fit is preferable.
In practice, one first determines the best fitting curves within the
two models. If the SSR of the single-band fit is smaller, this fit
is certainly preferred. However, if the two-band fit gives a smaller
SSR value, the F-test allows testing whether, within a fixed
confidence level (usually 5 \%), the improvement in the fit is not
only due to the increase in the number of parameters from 3 to 7. In
the two bottom panels of \fref{fig:totale_single}, the best-fitting
single-band BTK curve is represented by a dashed line. In the most
ambiguous case ($x=0.32$), the SSR in the range $[-10, 10]$ mV
(excluding the noisy regions of the curve) is $5.0 \cdot 10^{-4}$
for the two-band fit and $1.1 \cdot 10^{-3}$ for the single-band
one. The F-test shows that, for \textit{any} level of confidence,
the two-band fit is preferable. Another example, always in the
high-doping limit ($x=0.32$, $T\ped{c}\apex{A}=12.3$ K), is reported
in \fref{fig:parameters}. Here the best SSR in the range $[-10,10]$
mV is $1.75\cdot 10^{-3}$ in the single-band case and $9.67\cdot
10^{-4}$ in the two-band case. The F-test again shows that the
two-band fit is better for any level of confidence. The panels (b)
and (c) of the same figure give an idea of the spread of the fitting
parameters in different fits of the same curve; the range of
variation of the gaps is indicated by the grey strips. It is clear
in \fref{fig:parameters}(b) that $\Gamma_{\pi}$ is comparable or
equal to $\Delta_{\pi}$. This is a drawback of our experimental
technique, due to the aforementioned extrinsic broadening. It must
be said, however, that contrary to a widespread belief, the gap
structures are clearly visible in the conductance curves (and a
reliable gap measure can be extracted from their fit
\cite{gonnelli02c}) even if $\Delta=\Gamma$, provided that $Z$ is
not too small, as in our case. This can be easily shown by
calculating the conductance curves within the modified BTK model.

\appendix
\section*{Appendix B: Two-band Eliashberg equations}
\setcounter{section}{2}

Let us start from the generalization of the Eliashberg theory
\cite{eliashberg63} for systems with two bands, which has been
already used with success to study the MgB$_2$ system
\cite{brinkman02,choi02,golubov02}. To obtain the gaps and the
critical temperature within the s-wave, two-band Eliashberg model
one has to solve four coupled integral equations for the gaps
$\Delta_{i}(\mathrm{i}\omega_{n})$ and the renormalization functions
$Z_{i}(\mathrm{i}\omega_{n})$, where $i=\sigma, \pi$ is the band
index and $\omega_{n}$ are the Matsubara frequencies. We included in
the equations the non-magnetic impurity scattering rates in the Born
approximation, $\gamma_{ij}$.
 \begin{eqnarray}
\omega_{n}Z_{i}(\mathrm{i}\omega_{n})&=&\omega_{n}+\pi
T\sum_{m,j}\Lambda_{ij}(\mathrm{i}\omega_{n}-\mathrm{i}\omega_{m})N^{j}_{Z}(\mathrm{i}\omega_{m})+\nonumber\\
& &
+\sum_{j}\gamma_{ij}N^{j}_{Z}(\mathrm{i}\omega_{n})\label{eq:EE1}
\end{eqnarray}
\begin{eqnarray}
Z_{i}(\mathrm{i}\omega_{n})\Delta_{i}(\mathrm{i}\omega_{n})&=&\pi
T\sum_{m,j}[\Lambda_{ij}(\mathrm{i}\omega_{n}-\mathrm{i}\omega_{m})-\mu^{*}_{ij}(\omega_{c})]\cdot\nonumber\\
& &
\hspace{-2.5cm}\cdot\theta(|\omega_{c}|-\omega_{m})N^{j}_{\Delta}(\mathrm{i}\omega_{m})+\sum_{j}%
\gamma_{ij}N^{j}_{\Delta}(\mathrm{i}\omega_{n}) \label{eq:EE2}
\end{eqnarray}
where $\theta$ is the Heaviside function, $\omega_{c}$ is a cut-off
energy and
$\Lambda_{ij}(\mathrm{i}\omega_{n}-\mathrm{i}\omega_{m})=\int_{0}^{+\infty}d\omega
\alpha^{2}_{ij}(\omega)F(\omega)/[(\omega_{n}-\omega_{m})^{2}+\omega^{2}]$,
$N^{j}_{\Delta}(\mathrm{i}\omega_{m})=\Delta_{j}(\mathrm{i}\omega_{m})/
{\sqrt{\omega^{2}_{m}+\Delta^{2}_{j}(\mathrm{i}\omega_{m})}}$,
$N^{j}_{Z}(\mathrm{i}\omega_{m})=\omega_{m}/
{\sqrt{\omega^{2}_{m}+\Delta^{2}_{j}(\mathrm{i}\omega_{m})}}$.

The solution of the Eliashberg equations requires the following
input: i) four (but only three independent) electron-phonon spectral
functions $\alpha^{2}_{ij}(\omega)F(\omega)$; ii) four (but only
three independent) elements of the Coulomb pseudopotential matrix
$\mu^{*}(\omega\ped{c})$; iii) four (but only three independent)
impurity scattering rates $\gamma_{ij}$.

The four spectral functions $\alpha^{2}_{ij}(\omega)F(\omega)$ were
calculated for pure MgB$_2$ in ref. \cite{golubov02}. For
simplicity, we will assume here that the shape of the
$\alpha^{2}_{ij}(\omega)F(\omega,x)$ functions does not change with
$x$, and we will only scale their amplitude with the electron-phonon
coupling constants $\lambda_{ij}$:
\begin{equation}
\alpha_{ij}^2 F(\omega, x)=
\frac{\lambda_{ij}(x)}{\lambda_{ij}(x=0)}\alpha_{ij}^2 F(\omega,
x=0)
\end{equation}
where

\begin{equation}
\lambda_{ij}(x)=\frac{N_{N}^{j}(E_{F},x)}{N_{N}^{j}(E_{F},x=0)}\lambda_{ij}(x=0).
\end{equation}

As far as the Coulomb pseudopotential is concerned, we used the
expression calculated for pure MgB$_2$ \cite{mazin04}, though
including the dependence of the densities of states at the Fermi
level $N_{N}^{i}(E\ped{F},x)$ on the doping content $x$:
\begin{eqnarray}
\mu^{*}(x)& = & \left| \begin{array}{cc}%
\mu^{*}(x)\ped{\sigma \sigma} & \mu^{*}(x)\ped{\sigma \pi}\\
\mu^{*}(x)\ped{\pi \sigma} & \mu^{*}(x)\ped{\pi \pi}
\end{array} \right| =  \nonumber \\
& = & \mu_{0}N\ped{N}\apex{tot}(E\ped{F})
\left| \begin{array}{cc}%
\frac{2.23}{N\ped{N}\apex{\sigma}(E\ped{F},x)} &
\frac{1}{N\ped{N}\apex{\sigma}(E\ped{F},x)}\\ & \\
\frac{1}{N\ped{N}\apex{\pi}(E\ped{F},x)} &
\frac{2.48}{N\ped{N}\apex{\pi}(E\ped{F},x)}
\end{array} \right| \label{eq:mu}
\end{eqnarray}

As for the scattering rates, let us remind here that, according to
Anderson's theorem, the intraband scattering parameters
$\gamma_{ii}$ have no effect on either $T_{c}$ or the gaps so that
they can be dropped. We are thus left only with the interband
scattering rates $\gamma_{\sigma \pi}$ and $\gamma_{\pi \sigma}$,
which are however related through the equation
\begin{equation}
\gamma_{\pi \sigma}= \gamma_{\sigma \pi}
\frac{N_{\sigma}(E_{F})}{N_{\pi}(E_{F})}
\end{equation}

This is the reason why we can choose $\gamma_{\sigma \pi}$ as the
only adjustable parameter. Moreover, it can be shown that
\emph{only} the interband scattering can make the small gap increase
while both T$_c$ and the large gap decrease (on increasing x), in
the way experimentally observed.

It is worth clarifying the relationship between the broadening
parameters $\Gamma_{\sigma}$ and $\Gamma_{\pi}$ in the BTK model and
the scattering parameters $\gamma_{\sigma\sigma}$,
$\gamma_{\pi\pi}$, $\gamma_{\sigma \pi}$ in the Eliashberg theory.
The intrinsic BTK linewidth for a given band, e.g. $\Gamma_{\pi}$,
takes into account all the scattering channels and is thus
proportional to $\gamma_{\pi \pi}+\gamma_{\pi \sigma}$. A direct
relationship between the intrinsic $\Gamma_{\pi}$ and
$\gamma_{\sigma \pi}$ cannot be established unless an independent
determination of the intraband scattering rates is obtained.
Furthermore, in our case the values of $\Gamma_{\pi}$ extracted from
the fit contain an ``extrinsic'' term (related to inelastic
scattering in the vicinity of the interface) in addition to the
intrinsic linewidth. This prevents a direct connection between the
$\Gamma_{i}$ parameters of the BTK model and the $\gamma_{ij}$ of
the Eliashberg theory.

\appendix
\section*{Appendix C: A test of consistency}
\setcounter{section}{3}

With reference to \fref{fig:Eliashberg}, it is worth noting that we
are fitting with the Eliashberg theory -- including interband
scattering -- data points that were obtained through a BTK fit of
experimental curves (and thus without taking into account the
possible effect of the interband scattering on the curves
themselves, which is theoretically expected not to be negligible
\cite{bascones01}). This approximation is hardly avoidable since a
fit of the conductance curves within the two-band Eliashberg theory
would be a very complex (if possible) task. For the same reason, a
direct proof of the reasonability of the BTK approach -- that would
be obtained by directly comparing the results of the two fitting
procedures, BTK and Eliashberg -- cannot be obtained. However, one
can try to demonstrate that a theoretical curve calculated within
the two-band Eliashberg theory (with given values of the gaps and a
non-zero interband scattering rate) can be fitted with the BTK
model, and that the gap amplitudes resulting from the fit are
consistent with the original ones.

\begin{figure}[tb]
\begin{center}
\includegraphics[keepaspectratio, width=0.5\columnwidth]{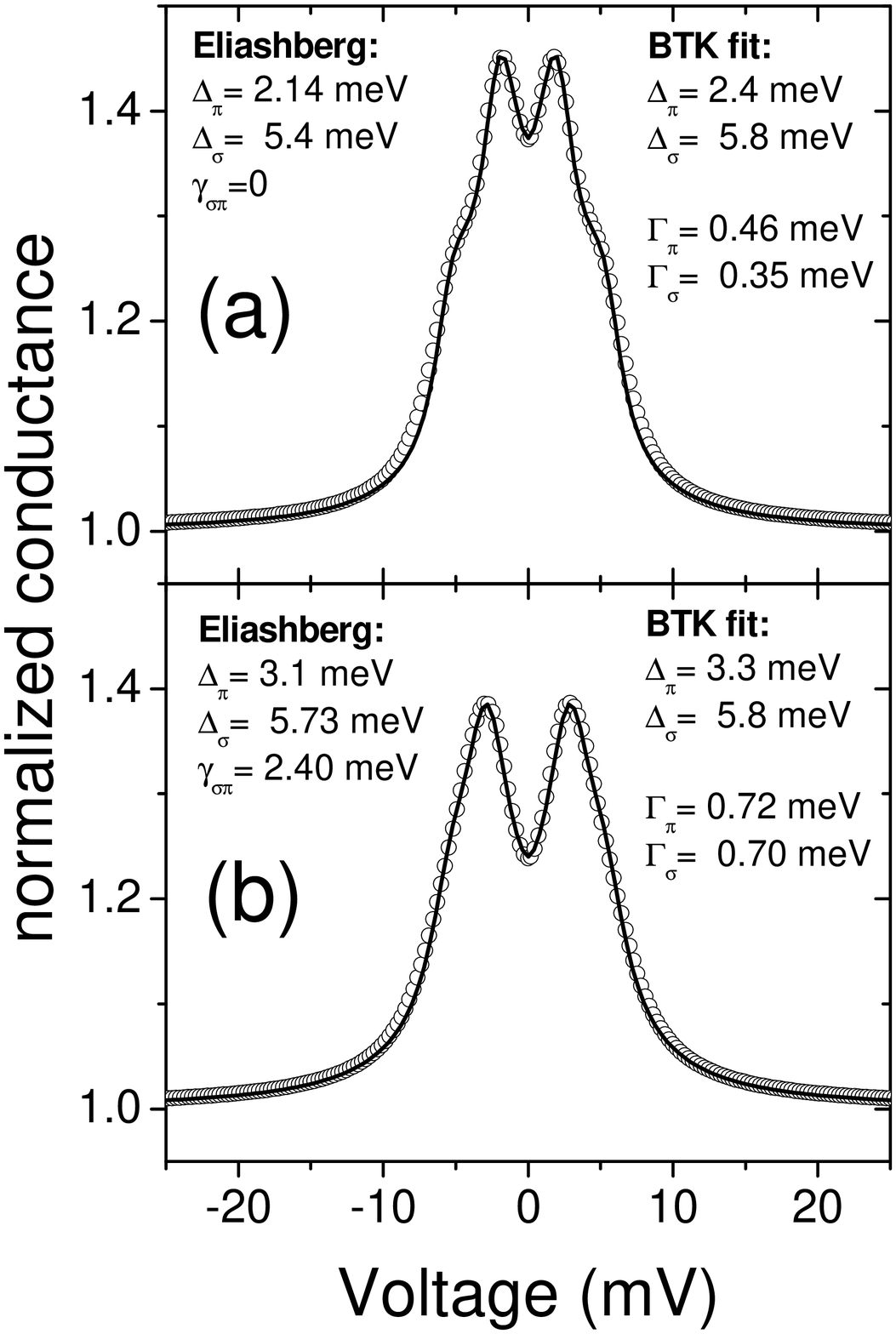}
\end{center}
\caption{\small{(a) Symbols: the theoretical (normalized)
Andreev-reflection conductance curve generated within Eliashberg
theory by using the parameters corresponding to the point at
$T_{c}^{A}$ = 30 K on the dashed curves of \fref{fig:Eliashberg}.
The values of the parameters are indicated on the left. Solid line:
best fit of the curve with the two-band BTK model. The fitting
parameters are indicated on the right. (b) Same as in (a), but with
the parameters corresponding to the point at $T_{c}^{A}$ = 30 K on
the solid curves of \fref{fig:Eliashberg}.}}
\label{fig:EliashbergAR}
\end{figure}

Let us refer for clarity to the point where the $\pi$-band gap is
maximum (see \fref{fig:Eliashberg}), which is also the most critical
one. We first calculated the gap functions $\Delta_{\sigma}(\omega)$
and $\Delta_{\pi}(\omega)$ ($\omega$ being the energy) within the
Eliashberg theory, with the parameters corresponding to the two
curves in \fref{fig:Eliashberg} (dashed and solid lines), that means
in particular with $\gamma_{\sigma \pi}$=0 and $\gamma_{\sigma
\pi}$=2.40 meV, respectively. Then, we calculated the corresponding
Andreev-reflection conductance curves at 4.2 K with \textit{no
additional smearing}, and using the experimental values of
$Z_{\sigma}$ and $Z_{\pi}$ \cite{kashiwaya96}. The two curves are
reported as open circles in \fref{fig:EliashbergAR}(a) and (b),
respectively.

Finally, we fitted these curves with the two-band BTK model and
compared the values of the gaps given by the fit with those used to
generate the curves. Note that the amplitude of the theoretical
curves in \fref{fig:EliashbergAR} ($\gtrsim$ 1.4) is greater than
that of the experimental ones, which are further smeared by
extrinsic broadening factors (e.g. related to the specific
measurement technique we use \cite{tanaka03,chalsani07}). This makes
the test even stricter. The best-fitting BTK curves are shown in
\fref{fig:EliashbergAR}(a) and (b) as solid lines. It is clearly
seen that both the gaps are rather well re-obtained, with an error
which is of the same order of magnitude as the experimental
uncertainty. This shows that the trend of the small gap
$\Delta_{\pi}$ obtained by the BTK fit of the conductance curves is
not due to an artifact introduced by the BTK fit itself. In other
words, the enhancement of $\Delta_{\pi}$ at intermediate $T_{c}^{A}$
values is a real effect that would be obtained as well by fitting
the experimental conductance curves with the more appropriate
Eliashberg theory.

\newpage

\section*{References}
%

\bibliography{bibliografia}


\end{document}